\documentclass[letterpaper, 10 pt, conference]{ieeeconf}  

\IEEEoverridecommandlockouts                              

\usepackage[printwatermark]{xwatermark}
\usepackage{xcolor}
\usepackage{multirow}
\usepackage{amsmath}
\usepackage{epstopdf}
\usepackage{amsthm}
\usepackage{tikz}
\usepackage{siunitx}
\usepackage{booktabs}
\usepackage{amssymb}
\usepackage{graphicx} 
\usepackage{subfig}
\usepackage{graphicx}
\usepackage{pdfpages}
\newtheorem{theorem}{Theorem}
\newtheorem{proposition}[theorem]{Proposition}

\theoremstyle{definition}

\newcommand{\etal}{\textit{et al}.}

\newcommand{\eg}{\textit{e}.\textit{g}. }

\title{\LARGE \bf
Episodic Koopman Learning of Nonlinear Robot Dynamics 
with Application to Fast Multirotor Landing}

\author{Carl Folkestad$^*$, Daniel Pastor$^*$, and Joel W. Burdick\thanks{*Both authors contributed equally}\thanks{All authors are with the Division of Engineering and Applied Sciences, California Institute of Technology, Pasadena, CA 91125, USA \texttt{\{carl.folkestad,dpastorm,jburdick\}@caltech.edu}}}

\begin{document}
\maketitle
\thispagestyle{empty}
\pagestyle{empty}

\begin{abstract}
This paper presents a novel episodic method to learn a robot's nonlinear dynamics model and an increasingly optimal control sequence for a set of tasks.  The method is based on the {\em Koopman operator} approach to nonlinear dynamical systems analysis, which models the flow of {\em observables} in a function space, rather than a flow in a state space.  Practically, this method estimates a nonlinear diffeomorphism that lifts the dynamics to a higher dimensional space where they are linear.  Efficient Model Predictive Control methods can then be applied to the lifted model.  This approach allows for real time implementation in on-board hardware, with rigorous incorporation of both input and state constraints during learning. We demonstrate the method in a real-time implementation of fast multirotor landing, where the nonlinear ground effect is learned and used to improve landing speed and quality.     
\end{abstract}
\section{Introduction}


While modeling and identification (ID) techniques are well developed for some robotic mechanisms, such as manipulators, there are an increasing number of applications where modeling and ID are difficult.  Consider a multirotor drone, whose basic flight mechanics in open air are well understood \cite{QuadDynamicsTomlin,MahoneyNonlinear}.  However, when multirotors fly close to the ground or a wall, or inside a narrow tunnel (e.g., for non-invasive inspection), the unmodeled effects of the complex vehicle-air-environment interaction can substantially reduce the drone's path tracking accuracy, and perhaps its stability.  While  ground effect models can be incorporated, their accuracy is limited, and their parameters must still be estimated in a slow process. This particular problem motivates this paper.  There are many other applications, such as soft robotic structures or robotic manipulation of soft materials, where first principles modeling and parameter identification remain challenging in practice.  Moreover, one must still design a controller for the nonlinear mechanics that are identified.

{\em Learning} can capture the salient aspects of a robot's complex mechanics and environmental interactions.  Gaussian process dynamical systems models \cite{GaussianDynamic} can identify nonlinear affine control models in a non-parametric way. Yet, effective nonlinear control design after identifying the model can be challenging. Model-free reinforcement learning (MFRL) \cite{Duan2016} learns feedback policies that implicitly incorporate the robot's dynamics.  However, sample efficiency is very low.  Moreover, while safety during MFRL is now possible \cite{Garcia2015, Cheng:AAAI2019}, one cannot yet guarantee that learned policies will satisfy performance requirements or state and actuator limits. 
\begin{figure}[t]
	\centering
	\includegraphics[width=0.32\linewidth]{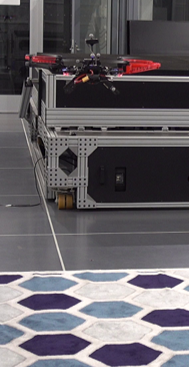}
	\includegraphics[width=0.32\linewidth]{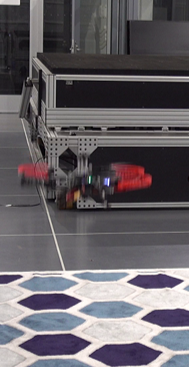}
	\includegraphics[width=0.32\linewidth]{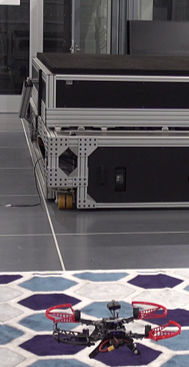}
	\caption{From left to right: hovering before the sequence start, high speed descent with learned dynamics, and soft landing.}
	\label{fig:car_video_frames}
	\vspace{-0.25 true in}
\end{figure}

This paper presents a new method, based on {\em Koopman spectral} analysis, to learn nonlinear robot dynamical models. Conventionally, a  system's behavior is characterized via its state space flows. In contrast, Koopman-based approaches study the evolution of {\em observables}, which are functions over the state-space.  In this space, the system can be represented by a {\em linear} (but possibly infinite dimensional) operator \cite{Lan2013, mauroy2016linear}.  Practically, the nonlinear dynamics are {\em lifted} to a higher dimensional space where they are linear. Efficient linear and optimal control principles can be applied to the lifted system.

Koopman inspired modeling and identification techniques have received substantial recent attention \cite{Budisic2012}.   The Dynamic Mode Decomposition (DMD) and extended DMD (EDMD) methods \cite{Taira2017} have been successfully used in the field of fluid mechanics to capture low-dimensional structure in complex flows.  More recently, Koopman-style modeling has been extended to {\em controlled} nonlinear systems \cite{Kaiser2017,Proctor2018}.

In practice, to model and identify a nonlinear system in the Koopman framework, one can identify a finite dimensional approximation to the linear Koopman operator, or identify the Koopman \textit{eigenfunctions}.  We use the latter approach (see Section \ref{sec:prelim_eigfunc}).  Previous methods for identifying Koopman eigenfunctions (e.g., \cite{Mohr2014,Korda2018}) depend upon assumptions that are problematical for robotic systems: the ID data is gathered while the robot operates under open loop controls, which can lead to catastrophic system damage.

In very recent work \cite{ACC2020}, the authors and collaborators have developed a {\em Koopman Eigenfunction Extended Dynamic Mode Decomposition} (KEEDMD) method to identify/learn an unknown nonlinear dynamical system from a batch of data gathered while the (robotic) system operates under a stabilizing (but not necessarily optimal) {\em linear controller}.

This paper substantially extends our prior work \cite{ACC2020} to make it practically useful for robotics.  First, the method gathers data while the system operates under any {\em nonlinear} stabilizing controller.  This enables input vector field nonlinearities to be captured, unlike prior Koopman-based model ID approaches. Second, we introduce an episodic learning procedure, by considering the closed-loop dynamics obtained with a non-linear controller as the autonomous dynamics for the next episode. This feature increases sample efficiency (i.e., fewer learning trials) for improving specific tasks, and enables nonlinear actuation effects, which are important in robotics, to be captured in the Koopman eigenfunctions. Third, it should be noted that data collected from robots executing trajectories formally violates the i.i.d. assumption underlying the performance guarantees of most learning paradigms. In practice, this fact can lead to error cascades and poor performance guarantees. Episodic learning mitigates this problem \cite{RossDagger}. 
Finally, our method integrates Model Predictive Control (MPC) \cite{Mayne2000} into its structure, thereby allowing control and state constraints to be satisfied during the learning process. 

Previous approaches have implemented MPC in real time on modest computational hardware \cite{Nicole2011}, on multirotors using an explicit solution in simulation \cite{liu2015explicit}, and designed feedback linearizing controllers for multirotors in real experiments~\cite{bangura2014real,abdolhosseini2013efficient}. Bouffard \etal~\cite{bouffard2012learning} also used MPC to learn ground effects using an experimental multirotor, but used an Extended Kalman Filter (EKF) in combination with Learning-Based MPC (LBMPC). Shi \etal \cite{Shi2018} experimentally demonstrated using a spectrally-normalized neural network to learn the ground effect and improve drone landing by designing a feedback linearizing controller utilizing the learned model. We introduce a new approach to solve this problem and aim to demonstrate that our method represents a first step towards practical Koopman-based learning and control of real-world robotic systems.


Section \ref{sec:preliminaries} reviews relevant facts about the Koopman operator and KEEDMD \cite{ACC2020}. Sections \ref{sec:episodic} and \ref{sec:EpisodicKEEDMD} introduce and develop the first main paper contribution--the episodic KEEDMD algorithm.  Section \ref{sec:multirotor} presents our second contribution, the experimental demonstration that our method can learn the ground effect in a fast landing multirotor.  

\section{Preliminaries on the Koopman Eigenfunction Extended Dynamic Mode Decomposition}\label{sec:preliminaries}

%
%

\subsection{The Koopman Operator}
Consider the autonomous dynamical system:
\begin{equation}\label{eq:aut_dynamics}
    \mathbf{\dot{x}} = f(\mathbf{x}) = A\mathbf{x} + v(\mathbf{x})
\end{equation}
with the state $x \in \mathcal{X} \subset \mathbb{R}^n$ and $f$ Lipschitz continuous on $\mathcal{X}$. We assume that the system (\ref{eq:aut_dynamics}) has a fixed point at the origin, $f(0) = 0$. The flow of this dynamical system is denoted by $S_t(x)$ and is defined as
\begin{equation}
    \frac{d}{dt}S_t(x) = f(S_t(x))
\end{equation}
for all $x \in \mathcal{X}$ and all $t\geq 0$. The \textit{Koopman operator semi-group} $(U_t)_{t\geq0}$, from now on simply denoted as the \textit{Koopman operator}, is defined as 
\begin{equation}
    U_tg = g \circ S_t
\end{equation}
for all $g \in \mathcal{C}(\mathcal{X})$, where $\circ$ denotes the function composition and $\mathcal{C}(\mathcal{X})$ is the space of all continuous functions on $\mathcal{X}$. Each element of the Koopman operator maps continuous functions to continuous functions, $U_t: \mathcal{C}(\mathcal{X}) \rightarrow \mathcal{C}(\mathcal{X})$. Crucially, each $U_t$ is a \textit{linear} operator. An \textit{eigenfunction} of the Koopman operator associated to an eigenvalue $e^{\lambda} \in \mathbb{C}$ is any function $\phi \in \mathcal{C}(\mathcal{X})$ that defines a coordinate evolving linearly along the flow of (\ref{eq:aut_dynamics}) satisfying
\begin{equation}
    (U_t \phi)(x) = \phi(S_t(x)) = e^{\lambda t} \phi (x).
\end{equation}
\subsection{Data-driven Construction of Koopman Eigenfunctions} \label{sec:prelim_eigfunc}

For any sufficiently smooth autonomous dynamical system that is asymptotically stable to a fixed point, Koopman eigenfunctions can be constructed by finding the eigenfunctions of the system's linearization around the fixed point and then composing them with a diffeomorphism \cite{Mohr2014}. The linearization of the dynamics (\ref{eq:aut_dynamics}) around the origin is

\begin{equation}\label{eq:lin_dynamics}
   \mathbf{\dot{y}} = \mathbf{D}f(0)\mathbf{y} = \hat{A}\mathbf{y}, \text{ } \mathbf{y} \in \mathcal{Y}
\end{equation}

The following proposition describes how to construct eigenfunction-eigenvalue pairs for the linearized system (\ref{eq:lin_dynamics}).
\begin{proposition} \label{prop:linear}
Let  $\hat{A}$ denote the linearization (\ref{eq:lin_dynamics}) of nonlinear system (\ref{eq:aut_dynamics}) and let $\{\mathbf{v}_1, \dots, \mathbf{v}_n\}$ be a basis of the eigenvectors of $\hat{A}$  corresponding to nonzero eigenvalues $\{\lambda_1, \dots, \lambda_n\}$. Let $\{\mathbf{w}_1, \dots, \mathbf{w}_n\}$ be an adjoint basis to $\{\mathbf{v}_1, \dots, \mathbf{v}_n\}$ such that $\langle \mathbf{v}_q,\mathbf{w}_r \rangle = \delta_{qr}$ and $\mathbf{w}_q$ is an eigenvector of $\hat{A}^*$ at eigenvalue $\Bar{\lambda}_q$. Then, the linear functional
\vskip -0.04 true in
\begin{equation} \label{eq:lin_eigfunc}
    \psi_q(\mathbf{y}) = \langle \mathbf{y}, \mathbf{w}_q \rangle
\end{equation}
\vskip -0.04 true in
\noindent is a nonzero eigenfunction of $U_{\hat{A}}$, the Koopman operator associated to $\hat{A}$. 
Furthermore, for any $(m_1, \dots, m_d) \in \mathbb{N}_0^d$
\vskip -0.04 true in
\begin{equation}\label{eq:eig_products}
    \bigg ( \prod_{q=1}^d e^{m_q\lambda_q}, \prod_{q=1}^d \psi_q^{m_q} \bigg )
\end{equation}{}
\vskip -0.04 true in
is an eigenpair of the Koopman operator $U_{\hat{A}}.$
\end{proposition}

These linear functionals (\ref{eq:lin_eigfunc}), termed \textit{principal eigenfunctions}, are used to construct the eigenfunctions associated with the Koopman operator of the nonlinear dynamics through the use of a {\em conjugacy map} (See \cite{Budisic2012}, Prop. 7).
\begin{proposition} \label{prop:conjugacy}
Assume that the nonlinear system (\ref{eq:aut_dynamics}) is topologically conjugate to the linearized system (\ref{eq:lin_dynamics}) via the diffeomorphism $h:\mathcal{X} \rightarrow \mathcal{Y}$. Let $B \in \mathcal{X}$ be a simply connected, bounded, positively invariant open set in $\mathcal{X}$ such that $h(B) \subset Q_r \subset \mathcal{Y}$, where $Q_r$ is a cube in $\mathcal{Y}$. Scaling $Q_r$ to the unit cube $Q_1$ via the smooth diffeomorphism $g: Q_r \rightarrow Q_1$ gives $(g \circ h)(B) \subset Q_1$. Then, if $\psi$ is an eigenfunction for $U_{\hat{A}}$ at $e^\lambda$, then $\psi \circ g \circ h$ is an eigenfunction for $U_f$ at eigenvalue $e^\lambda$, where $U_f$ is the Koopman operator associated with the nonlinear dynamics (\ref{eq:aut_dynamics}).
\end{proposition}
\noindent An  extension of the Hartman-Grobman theorem (\cite{Lan2013}, Theorem 2.3) guarantees the existence of a $\mathcal{C}^1$ diffeomorphism 
  \begin{equation}\label{eq:diffeo}
      y = c(\mathbf{x})=\mathbf{x}+h(\mathbf{x})
  \end{equation} 
between the linearized and nonlinear systems in the entire basin of attraction of a fixed point, such that $\mathbf{D}c(\mathbf{0})=I$.

\subsection{KEEDMD with Trajectory-tracking Control Laws}\label{sec:prelim_keedmd}
\begin{figure}[b]
\vskip -0.1 true in
\centering
\small
\begin{tabular}{p{8cm}}
\hline
\textbf{Algorithm 1} Data-driven Koopman Eigenpair Construction\\
\hline
\textbf{Require:} Data set $\mathcal{D} = \big((\mathbf{x}_k^j, \mathbf{u}_k^j)_{k=0}^{M_s}\big)_{j=1}^{M_t}$, nominal model matrices $A_{nom}$, $B_{nom}$, nominal control gains $K_{nom}$, desired trajectory $\boldsymbol{\tau}(t)$, number of lifting functions $N$, $N$ power combinations $(m_1^{(i)},\dots,m_d^{(i)})\in\mathbb{N}_0^d, i = 1,\dots,N$ \vspace{0.1cm}\\
Construct principal eigenpairs for the linearized dynamics: $(e^\lambda_j, \psi_j(\mathbf{y})) \leftarrow{} (e^\lambda_j, \langle \mathbf{y}, \mathbf{w}_j \rangle), \qquad j=1,\dots,n$\\
Construct $N$ eigenpairs from the principal eigenpairs:\\ $(\Tilde{e}^\lambda_i, \Tilde{\psi}_i) \leftarrow \big ( \prod_{j=1}^d e^{m_j^{(i)} \lambda_j}, \prod_{j=1}^d \psi_j^{m_j^{(i)}} \big ),\quad i=1,\dots,N$\\
Fit diffeomorphism estimator: $h(\mathbf{y}) \leftarrow \text{ERM}(\mathcal{H}_h, \mathcal{L}_h, \mathcal{D})$\\
Construct scaling function: $g(\mathbf{y}) \leftarrow g: \mathcal{Q}_r \rightarrow \mathcal{Q}_1$\\
Construct $N$ eigenpairs for the nonlinear dynamics:\\ $(\Tilde{e}^\lambda_i, \phi_i) \leftarrow (\Tilde{e}^\lambda_i, \Tilde{\psi}_i(g(h(\mathbf{y})))), \qquad i=1,\dots,N$ \vspace{0.1cm}\\
\textbf{Output:} $\Lambda = \text{diag}(\Tilde{e}^\lambda_1,\dots,\Tilde{e}^\lambda_N), \qquad \boldsymbol{\phi} = [\phi_1,\dots,\phi_N]^T$\\
     \hline
\end{tabular}
\end{figure}

\noindent Data-driven construction of Koopman eigenfunctions for nonlinear dynamics of the form $\dot{\mathbf{x}} = a(\mathbf{x}) + B\mathbf{u}$ is based on Section \ref{sec:prelim_eigfunc}, and summarized in Algorithm 1. General nonlinear dynamics will be considered in Section \ref{sec:episodic}.  For trajectory-tracking control laws, the algorithm assumes that a linearized nominal dynamics model $\mathbf{\dot{y}} = A_{nom}\mathbf{y} + B_{nom}\mathbf{u}$ is known, along with a nominal stabilizing feedback control law $\mathbf{u}(t) = K_{nom}(\mathbf{y}(t) - \boldsymbol{\tau}(t))$, where $\boldsymbol{\tau}(t) \in \mathcal{X}$ is the trajectory we want the system to track. Typically, $A_{nom},B_{nom}$ is the linearization of the dynamics around the origin and $K_{nom}$ the controller gains determined by e.g. LQR on the nominal state space model. Principal eigenfunctions, with eigenvalues $e^\lambda_j$, of the Koopman operator for the closed loop linearized dynamics $\mathbf{\dot{y}} = (A_{nom} + B_{nom}K_{nom}) \mathbf{y}$ are constructed using Prop. \ref{prop:linear}: $\psi_j(\mathbf{y}) = \langle \mathbf{y}, \mathbf{w}\rangle$, where $\mathbf{w}$ is an adjoint basis of the eigenvectors of $A_{cl} = (A_{nom}+B_{nom}K_{nom})$.  Products and powers (\ref{eq:eig_products}) generate arbitrarily many eigenpairs of the linearized system before applying the diffeomorphism (\ref{eq:diffeo}) between the nonlinear and linearized dynamics. This diffeomorphism $h: \mathcal{X} \rightarrow \mathcal{Y}$ is learned in a supervised way (\eg a neural network trained with gradient descent) by performing empirical risk minimization (ERM) of an appropriate loss function over a model class $\mathcal{H}_h$. For the trajectory-tracking case, the loss function is of the form
\begin{align}\label{eq:diff_loss}
\begin{split}
    &\mathcal{L}_h(\mathbf{x},\mathbf{\dot{x}}, A_{cl}\mathbf{x}-\mathbf{\dot{x}}, \boldsymbol{\tau}(t)) = \\
    & \quad \, || \dot{h}(\mathbf{x}) - A_{cl}h(\mathbf{x}) - (A_{cl}\mathbf{x} - \mathbf{\dot{x}}) + B_{nom}K_{nom}\boldsymbol{\tau}||^2 \\
\end{split}
\end{align}
(see \cite{ACC2020} for details). Finally, a function scaling $\mathcal{Y}\subset \mathcal{Q}_r$ into $Q_1$, where $\mathcal{Q}_r$ is a hyper cube of the same dimension as $\mathcal{Y}$ with radius $r$, is constructed (i.e. by scaling each coordinate into a unit cube) and approximate Koopman eigenpairs for the unknown, nonlinear dynamics are constructed: $(e^{\lambda_i}, \phi_i) = (\Tilde{e}^\lambda_i, \Tilde{\psi}_i(g(h(\mathbf{y}))))$, where $\Tilde{e}^\lambda_i, \Tilde{\psi}_i$ are the eigenpairs constructed with (\ref{eq:eig_products}).

KEEDMD uses the constructed eigenfunctions to lift the system states to a higher dimensional space where a linear dynamical model of the form $\mathbf{\dot{z}} = A\mathbf{z} + B\mathbf{u}$ can be identified. For Lagrangian dynamics, as the example in Section~\ref{sec:multirotor}, we use $\mathbf{z} = [\mathbf{x}, \boldsymbol{\phi}(\mathbf{x})]^T$ as the lifted state, where $\mathbf{x} = [\mathbf{p} \quad \mathbf{v}]^T$, $\textbf{p}$ the position, $\dot{\boldsymbol{p}}=\textbf{v}$ the velocity, and $\boldsymbol{\phi}$ is a vector of the eigenfunctions. While helping data efficiency, the method does not generally require any \textit{a priori} information of the structure of the dynamics. Since the time evolution of the eigenfunctions is dictated by their eigenvalues, we can show that the lifted state space model has the structure
\begin{equation*}\label{eq:KEEDMD}
\begin{bmatrix} \mathbf{\dot{p}} \\ \mathbf{\dot{v}} \\ \boldsymbol{\dot{\phi}} \bigg(\!\!\begin{bmatrix} \mathbf{p} \\ \mathbf{v} \end{bmatrix}\!\!\bigg) \end{bmatrix} =\underbrace{\begin{bmatrix} \begin{array}{cc} 0 & \quad I \end{array}{} & 0\\
\begin{array}{cc} A_{\mathbf{v}\mathbf{p}}& A_{\mathbf{v}\mathbf{v}} \end{array}{}&A_{\mathbf{v}\boldsymbol{\phi}} \\
-B_{\boldsymbol{\phi}}K_{nom} & \begin{array}{c}  \\  \end{array}{} \Lambda \begin{array}{c} \\ \end{array}{} \end{bmatrix}}_A
\begin{bmatrix} \mathbf{p} \\ \mathbf{v} \\ \boldsymbol{\phi} \bigg(\!\!\begin{bmatrix} \mathbf{p} \\ \mathbf{v} \end{bmatrix}\!\!\bigg) \end{bmatrix}
+\underbrace{\begin{bmatrix} B_\mathbf{p} \\
B_\mathbf{v}\\
B_{\boldsymbol{\phi}} \begin{array}{c}  \\  \end{array}{} \end{bmatrix}}_B  \mathbf{u}
\end{equation*}
where $0, I, \Lambda, K_{nom}$ are fixed matrices and $A_{\mathbf{v}\mathbf{p}}, A_{\mathbf{v}\mathbf{v}}$, $A_{\mathbf{v}\boldsymbol{\phi}}$, $B_{\mathbf{p}}$, $B_{\mathbf{v}}$, $B_{\boldsymbol{\phi}}$ are determined from data, and the term $-B_{\boldsymbol{\phi}} K_{nom}$ accounts for the nominal controller's state feedback effect on the evolution of the eigenfunctions. The desired trajectory effect of the controller is captured by the learning framework. The unknown elements of $A$ and $B$ can be found using linear regression on the data collected under the nominal control law (see \cite{ACC2020} for details). We define $\mathcal{L}_z$ as the mean squared error for the regression problem, where different forms of regularization can be included if needed. As the feedback controller is state-dependent, it is not possible to disambiguate its effect from the passive uncontrolled system dynamics in the learning process. To avoid an ill-conditioned KEEDMD regression problem, Brownian noise is added to perturb the nominal controller \cite{Kaiser}. Brownian noise is chosen in this instance because pure sampling from e.g. a Gaussian distribution leads to perturbations that have too high frequency to perturb the movement of the multirotor. This perturbation is also used by our episodic learning framework (Section \ref{sec:EpisodicKEEDMD}). When the lifted state space model is identified, state estimates can be obtained as $\mathbf{x} = C \mathbf{z}$, where $C = [I \quad 0]$. $C$ is denoted the \textit{projection matrix} of the lifted state space model.

\section{Episodic KEEDMD learning }\label{sec:episodic}
\begin{figure*}[t] 
    \centering
        \vskip 0.1 true in
        \includegraphics[width=\linewidth]{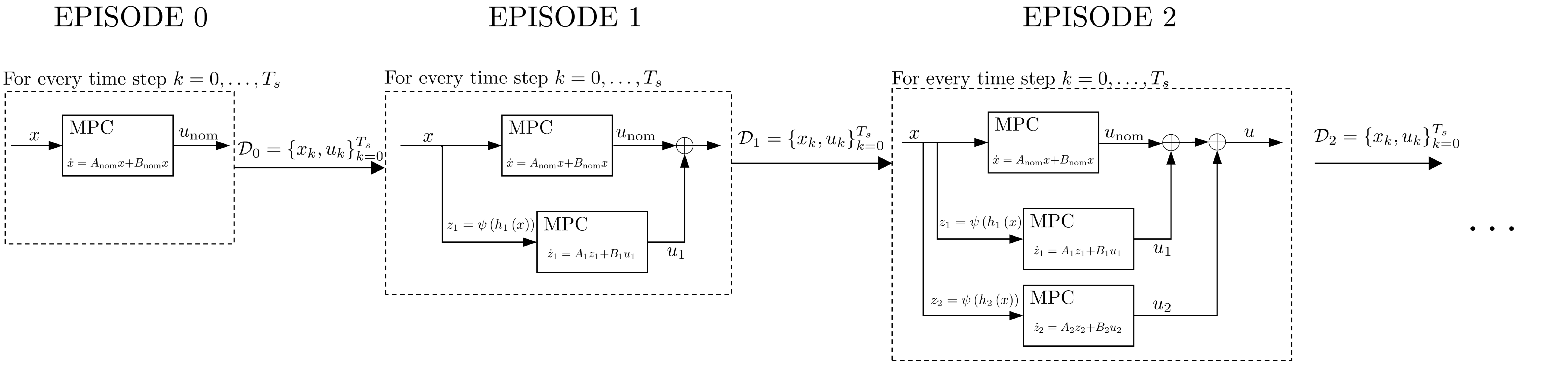} 
        \caption{Flow chart showing the different elements for each episode.}
    \label{fig:diagram}
    \vskip -0.3 true in
\end{figure*}
\subsection{Problem Setup and Dynamics Modeling}
Assume that we have selected a fixed trajectory $\boldsymbol{\tau}$ to be tracked by the robot during episodic learning. Further assume a nominal controller $\mathbf{\hat{u}}(\mathbf{x},\boldsymbol{\tau},t)$ that can stabilize the system to $\boldsymbol{\tau}$ within a region of attraction $\Omega$ around the trajectory.  This controller might be the outcome of a previous learning episode (see below), or the simple linear nominal controller from KEEDMD (Section \ref{sec:preliminaries}). Finally, the system's governing dynamics are assumed to be {\em unknown} 
\begin{equation}\label{eq:true_dyn}
    \mathbf{\dot{x}} = f(\mathbf{x},\mathbf{u})
\end{equation}
where $\mathbf{x} \in \mathcal{X} \subset \mathbb{R}^n, u \in \mathcal{U} \subset \mathbb{R}^m$. and $f(\mathbf{x},\mathbf{u})$ is assumed to be Lipschitz continuous on $\mathcal{X} \times \mathcal{U}$. 


\subsection{Learning with Arbitrary Stabilizing Control Laws} \label{sec:KEEDMD_extension}

KEEDMD (Section \ref{sec:preliminaries}) requires batch training data to be collected from system executions under a nominal linear control law: $\mathbf{u}_{nom}(\mathbf{x}) = K_{nom}(\mathbf{x} - \boldsymbol{\tau}(t))$. A main contribution of this paper is to iteratively learn an improving sequence of eigenfunctions and nonlinear controllers. Specifically, we will iteratively use the lifted state-space model to design an MPC-controller to track learning trajectories (see Figure~\ref{fig:diagram}). 

If a candidate nonlinear controller $\mathbf{\hat{u}}(\mathbf{x}, \boldsymbol{\tau},t)$ can stabilize system (\ref{eq:true_dyn}) to a given trajectory $\boldsymbol{\tau}$, the controlled system can be described by the autonomous dynamics
  \begin{equation}\label{eq:aut_dyn_ctrl}
       \mathbf{\dot{x}} = f(\mathbf{x},\mathbf{\hat{u}}(\mathbf{x},\boldsymbol{\tau},t)) = F_{\mathbf{\hat{u}},\boldsymbol{\tau}}(\mathbf{x},t) .
  \end{equation}
Importantly, for the autonomous dynamics (\ref{eq:aut_dyn_ctrl}), there exists an associated Koopman operator $U_{F_{\mathbf{\hat{u}},\boldsymbol{\tau}}}$ that depends on control law $\mathbf{\hat{u}}$ and trajectory $\boldsymbol{\tau}$.  Therefore, approximate eigenpairs for $U_{F_{\mathbf{\hat{u}},\boldsymbol{\tau}}}$ can be constructed (see Section \ref{sec:prelim_eigfunc}) from the gathered state and control samples. A lifted state-space model can be constructed from these eigenpairs. 

However, unlike the framework reviewed in Section \ref{sec:preliminaries}, we aim to learn a dynamical model that assumes that the system is already regulated by the \textit{nominal controller} $\mathbf{\hat{u}}(\mathbf{x}, \boldsymbol{\tau},t)$. As a result, the $A$-matrix of the lifted state space model captures the autonomous dynamics under the nominal control law (Eq. \ref{eq:aut_dyn_ctrl}), and the $B$-matrix captures the effect of control variations around the nominal controller:
\begin{equation}\label{eq:centered_model}
    \mathbf{\dot{\hat{z}}} = A\mathbf{\hat{z}} + B(\mathbf{u}(\mathbf{x},\boldsymbol{\tau},t) - \mathbf{\hat{u}}(\mathbf{x},\boldsymbol{\tau},t)).
\end{equation}
This model is used in an MPC framework below to design an \textit{augmenting} control law that adds optimal control actions to the nominal controller. The augmenting controller leverages the improved system model to make corrections to sub-optimal actions taken by the nominal controller.

\subsection{Modifications to Allow the Diffeomorphism to Capture Nonlinear Control and Dynamics Effects}
To enable the learning framework to capture nonlinear effects caused by the nonlinear controller and actuated dynamics, a minor modification to the function approximator of $h$ is necessary. Namely, since the diffeomorphism is affected by the forcing signal $\boldsymbol{\tau}(t)$ it must be included in the inputs of $h$. This is motivated by the form of the diffeomorphism loss function (\ref{eq:diff_loss}). In the case considered in the preliminaries however, the actuated dynamics and controller are assumed to be linear. This causes the effect of the forcing signal $\boldsymbol{\tau}(t)$ to cancel out such that the diffeomorphism is independent of the desired trajectory. In the general nonlinear case however, the effect is not canceled out and must be captured by the diffeomorphism. As a result, the diffeomorphism is modified such that $h: \mathcal{X} \times \mathcal{X} \rightarrow \mathcal{Y}$ (see \cite{ACC2020} for details).

\section{Episodic Eigenfunction Construction and KEEDMD Inference} \label{sec:EpisodicKEEDMD}
This section describes the main contribution of this paper, a substantial extension of the KEEDMD framework to allow iterative learning and improvement of the lifted state-space model and its associated controller. 

\subsection{Overview of the Episodic Learning Algorithm}
Algorithm 2 summarizes our episodic learning approach, which applies three key steps per episode. In each episode, $e$, the first key step starts when an initial condition is sampled from set $X_0$ and an experiment is executed with the controller that results from the previous episode $\mathbf{u}_{e-1}(\mathbf{x},\boldsymbol{\tau},t)$. The state $\mathbf{x}$, control actions $\mathbf{u}_{e-1}$, Brownian noise control perturbations $\mathbf{\Tilde{u}}$, and the desired position dictated by the trajectory at the time associated with the $i$-th sample $\boldsymbol{\tau}_i$ are sampled. State data can be differentiated numerically to find estimates $\mathbf{\dot{x}}$. The resulting data set is:
\begin{equation}
    \mathcal{D}_x^{(e)} = \left\{\left(\mathbf{x}_i^{(e)}, \mathbf{u}_i^{(e)}, \mathbf{\Tilde{u}}_i^{(e)}, \boldsymbol{\tau}_i\right), \mathbf{\dot{x}}_i^{(e)}\right\}_{i=1}^{T_s}
\end{equation}
where $\mathbf{x}_i^{(e)}$ denotes the $i$-th timestep of the $e$-th episode and $T_s$ denotes the number of samples in the episode. From $\mathcal{D}_x^{(e)}$ we estimate the diffeomorphism $h$ and construct the eigenfunctions $\boldsymbol{\phi}^{(e)}(\mathbf{x})$ with associated eigenvalues $\Lambda^{(e)}$, via Algorithm 1. Since changes in the control law between episodes are expected to be small, we warm start the learning algorithm with model coefficients from the previous episode. 
\begin{figure}[b]
\centering
\small
\begin{tabular}{p{8.1cm}}
     \hline
     \textbf{Algorithm 2} Episodic KEEDMD\\
     \hline
     \textbf{Require:} Desired trajectory $\boldsymbol{\tau}$, nominal controller $\mathbf{\hat{u}}(\mathbf{x},\boldsymbol{\tau},t)$, diffeomorphism model class $\mathcal{H}_h$, diffeomorphism loss $\mathcal{L}_h$, number of lifting functions $N$, KEEDMD loss $\mathcal{L}_z$ \vspace{0.1cm}\\
     $\mathcal{D}_z = \emptyset,\quad \mathbf{u}_0(\mathbf{x},\boldsymbol{\tau},t) =  \mathbf{\hat{u}}(\mathbf{x},\boldsymbol{\tau},t)$\\
     \textbf{for} $e=1,\dots,N_{ep}$ \textbf{do}\\
     $\quad$ Sample initial condition: $\mathbf{x_0} \leftarrow \text{sample}(X_0)$\\
     $\quad$ Execute experiment: $\mathcal{D}_x^{(e)} \leftarrow \text{run}(\mathbf{x}_0, \mathbf{u}^{(e-1)}(\mathbf{x},\boldsymbol{\tau},t))$\\
     $\quad$ Fit diffeomorphism estimator: $h(\mathbf{x}) \leftarrow \text{ERM}(\mathcal{H}_h, \mathcal{L}_h, \mathcal{D}_x^{(e)})$\\
     $\quad$ Construct eigenpairs: $(\boldsymbol{\phi}^{(e)}(\mathbf{x}), \Lambda^{(e)}) \leftarrow h(g(\boldsymbol{\psi}(\mathbf{x})))$\\ 
     $\quad$ Construct and aggregate lifted data set: $\mathcal{D}_z \leftarrow \mathcal{D}_z \cup \mathcal{D}_z^{(e)}$\\
     $\quad$ Fit KEEDMD model: $\mathbf{\dot{z}}^{(e)}(\mathbf{z}) \leftarrow \text{ERM}((\boldsymbol{\phi}^{(e)} \!,\! \Lambda^{(e)}), \mathcal{L}_z, \mathcal{D}_z)$ \\
     $\quad$ Update controller: $\mathbf{u}^{(e)} \leftarrow \mathbf{u}^{(e-1)} + w^{(e)}\text{MPC}(\mathbf{\dot{z}}^{(e)}, \mathbf{u}^{(e-1)})$\\
     \textbf{end for}\vspace{0.1cm}\\
     \textbf{Output:} Final control law $\mathbf{u}^{(N_{ep})}$\\
     \hline
\end{tabular}
\end{figure}

\begin{figure*}[t] 
    \vskip -0.15 true in
    \centering
    \subfloat{\includegraphics[height=7.2cm]{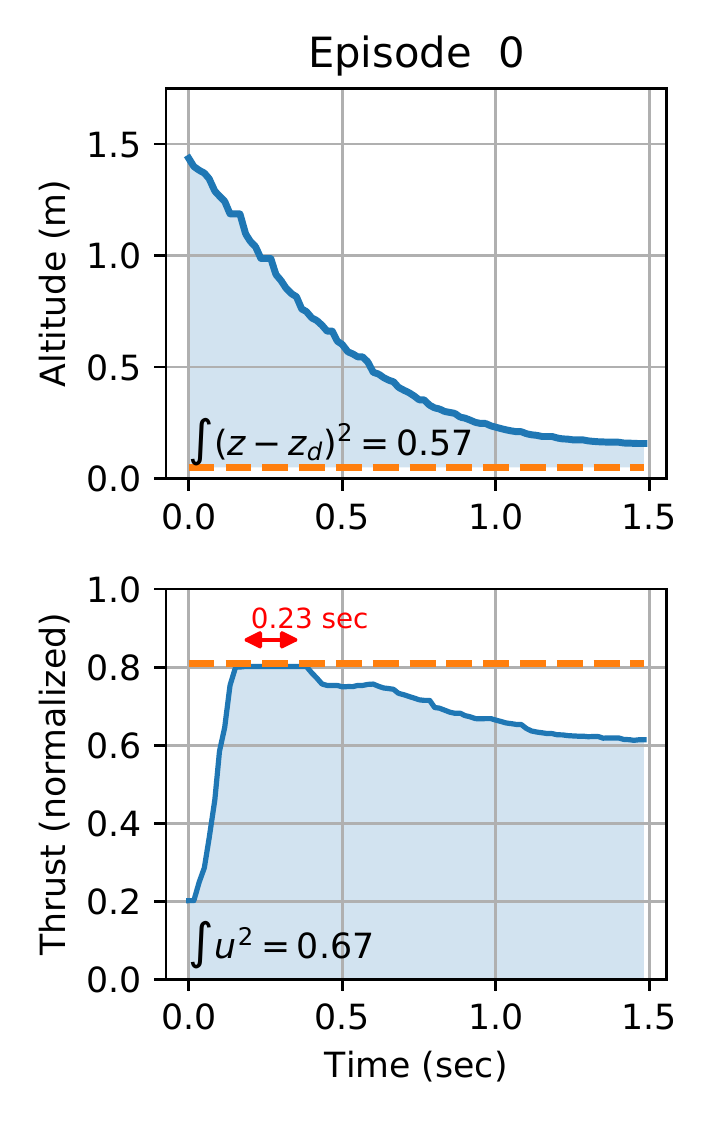}\label{fig:track1}} 
    \subfloat{\includegraphics[height=7.2cm]{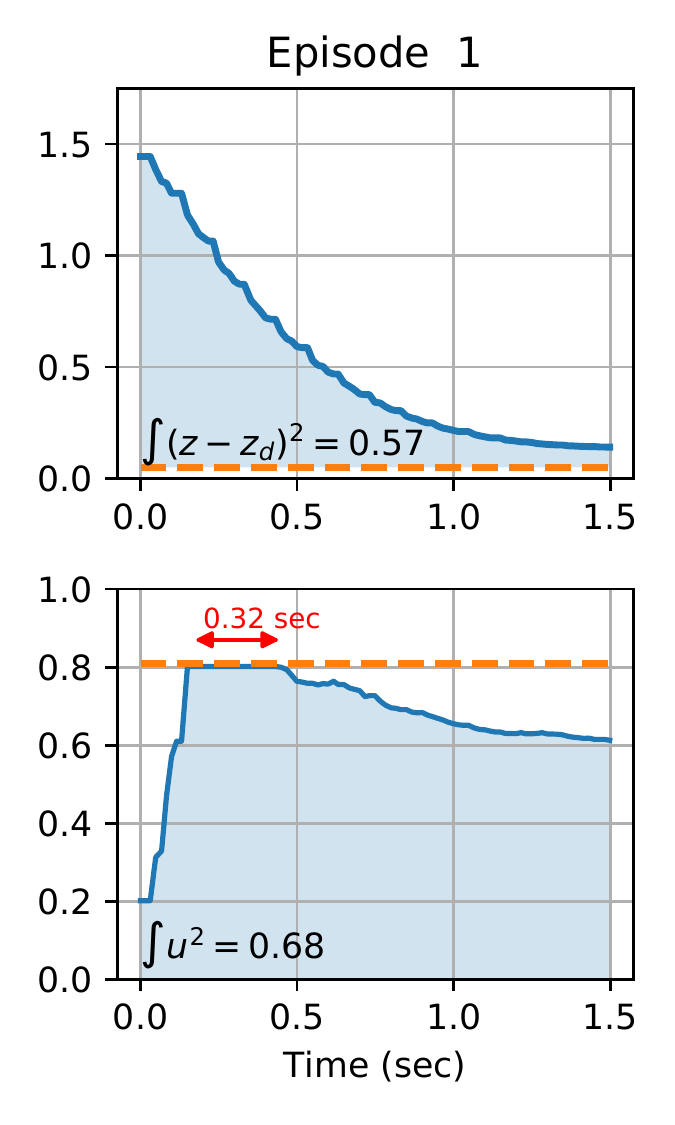}\label{fig:track2}} 
    \subfloat{\includegraphics[height=7.2cm]{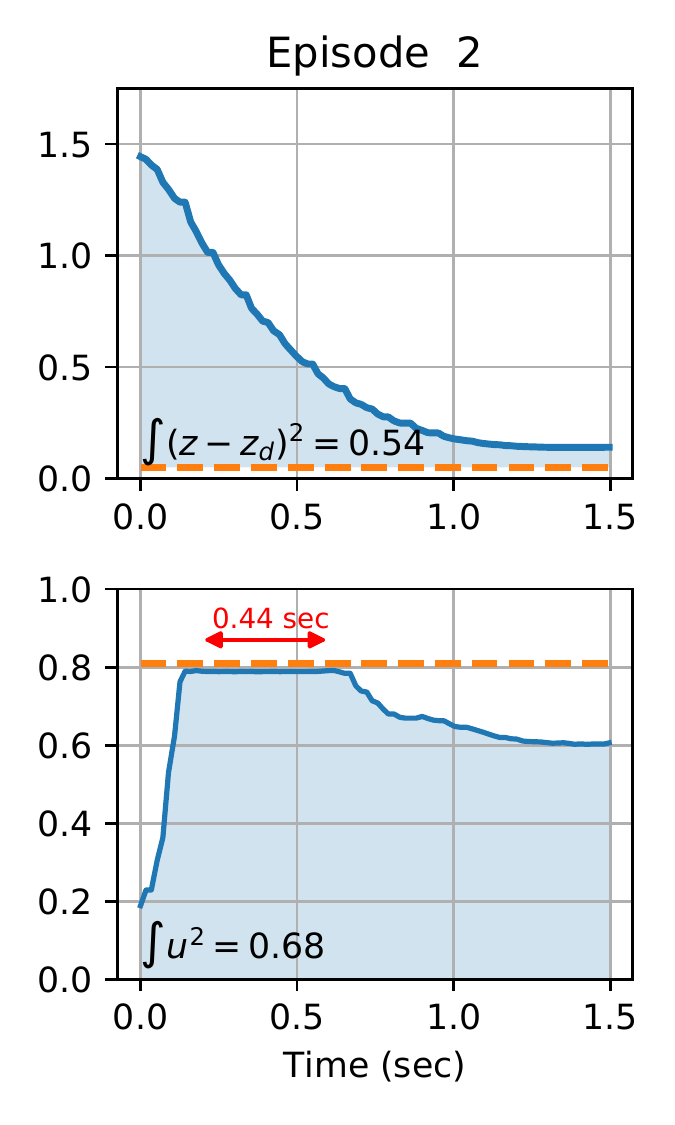}\label{fig:track3}} 
    \subfloat{\includegraphics[height=7.2cm]{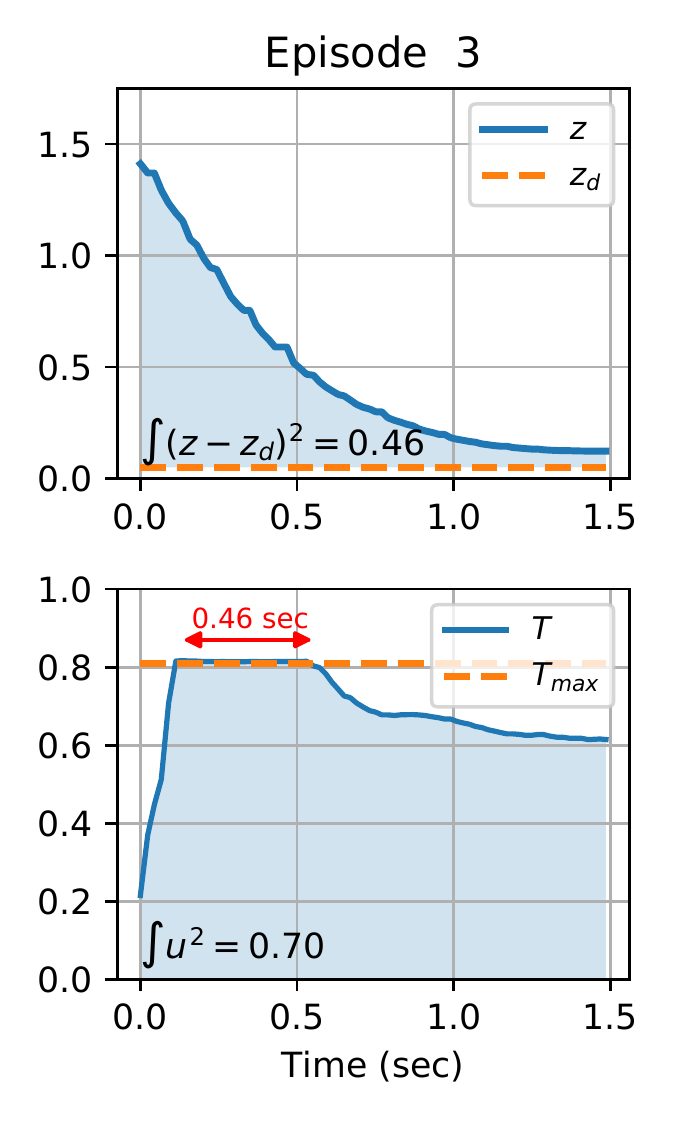}\label{fig:track4}} 
    \caption{Evolution of drone altitude $p_z$ with accumulated error and control effort after each episode. Episode 0: baseline controller, Episode 1-3: performance after each episode of learning. Red arrows: duration the thrust constraint is active.}
    \label{fig:ep_evolution}
    \vskip -0.2 true in
\end{figure*}
The second key step is to use the constructed eigenpairs to build a lifted data set $\mathcal{D}_z^{(e)}$. 
\begin{equation}
    \mathcal{D}_z^{(e)} = \left\{\left(\mathbf{z}_i^{(e)},\mathbf{u}_i^{(e)},\mathbf{\Tilde{u}}_i^{(e)}, \boldsymbol{\tau}_i\right),\mathbf{\dot{z}}_i^{(e)} \right\}_{i=1}^{T_s}
\end{equation}
which is the same data as $\mathcal{D}_x^{(e)}$, but with the state and its derivative, $\mathbf{x}_i^{(e)},\mathbf{\dot{x}}_i^{(e)}$, replaced with the lifted state and its derivative, $\mathbf{z}_i^{(e)}$, $\mathbf{\dot{z}}_i^{(e)}$. Next, data from the current and previous episodes is aggregated:  $\bigcup_{j=1}^e\mathcal{D}_z^{(j)}$. The lifted state-space model is constructed from this data using the framework of Section \ref{sec:KEEDMD_extension}. 
This results in a model of the form (\ref{eq:centered_model}).

In the third and final step, an augmenting MPC is designed (see Section \ref{sec:MPC_details}) for the lifted state-space model. The evaluation of the previous iteration's controllers is necessitated by the fact that the eigenfunctions depend on the dynamics under closed loop control with the controller deployed in the previous episodes. The controller augmentations are weighted and added to the previous episode's control law: $\mathbf{u}_e = \mathbf{u}_0 + \sum_{j=1}^{e}w_j\mathbf{u}_j$, where $w_e$ is a weighting factor indicating the confidence in the augmenting controller. The weighting factors can be any monotonically increasing sequence on the interval $[0,1]$ which allows the augmenting controller to have a bigger impact after a sufficiently rich data set has been collected. 

\subsection{Model Predictive Controller Details}\label{sec:MPC_details}

Inspired by \cite{Korda2018a}, we transform the original non-linear optimization problem into an efficient quadratic program (QP). The QP formulation requires us to discretize the previously learned linear continuous dynamics. We assume a known objective function of states and controls only. For simplicity, we use a quadratic objective function with respect to the state error and control action, but other objective functions can be used by simply adding it to the lifting functions. We assume known control bounds $u_{\min}, u_{\max} \in \mathbb{R}^{m}$ and state bounds $x_{\min}, x_{\max} \in \mathbb{R}^{n}$. Because the control input for each MPC problem refers to the change from the the previous controller, we have to correct for this change in the control bounds. All these assumptions define the following optimization problem that is solved at each time step:

\begin{equation}\label{eq:ZMPC}
\begin{array}{ll}{\underset{\substack{u \in \mathbb{R}^{m \times N_p} \\ z\in \mathbb{R}^{N\times N_p}}}{\operatorname{min}}} & {\sum_{p=1}^{N_p}\left[\left(C_jz_{p}-\tau_p\right)^{T} Q\left(C_jz_{p}-\tau_p\right)+u_{p}^{T} R u_{p}\right]} \\ {\text{s.t.}} & {z_{p}=A_jz_{p-1}+B_ju_{p}} \\ {} & {x_{\min } \leq C_jz_{p} \leq x_{\max }} \,\,\,\,\,\,\,\,\,\,\,\,\,\,\,\, p=1,\dots,N_p\\ {} & {u_{\min } \leq u_{p}-\sum_{i=1}^{j-1} w^{(i)}u_p^{(i)} \leq u_{\max }} \\ {} & {z_{0}=\phi_j(x_k)}\end{array}
\end{equation}

\noindent where $Q \in \mathbb{R}^{n \times n}$ and $ R \in \mathbb{R}^{m \times m}$ are positive semidefinite cost matrices, $\tau \in \mathbb{R}^{n \times N_p}$ is the reference trajectory, $A_j \in \mathbb{R}^{N\times N}$ and $B_j\in \mathbb{R}^{N\times m}$ are the discrete time versions of (\ref{eq:centered_model}) for controller $j$, $C_j \in \mathbb{R}^{n\times N}$ is the $j^{th}$ controller's projection matrix, and $\phi_j \in \mathbb{R}^{N}$ are the $j^{th}$ controller's eigenfunctions. See Figure~\ref{fig:diagram} to see how each controller is used as more episodes are being executed. In addition, we add a smoothing regularizer to avoid chatter that may arise from optimization-based controllers \cite{Morris2015} of the form $\sum_{p=1}^{N_p}\alpha_R(u_p-u_{p-1})^2$ where $u_0$ is the deployed control action at the previous timestep.



%
\section{Improving Fast multirotor Descent and Landing by Learning the Ground Effect}\label{sec:multirotor}

To validate our methodology, we apply it to fast descent and landing of a multirotor\footnote{The code for learning and control is publicly available on\\ \texttt{https://github.com/Cafolkes/keedmd}}. As the vehicle approaches the landing plane, a ground effect from the interaction of the prop downwash and the landing surface becomes prominent. This effect induces added upward thrust on the drone, which can lead to poor tracking performance for control designs that rely on models which omit these fluid flow interactions.

\subsection{Modeling and Problem Statement}

To simplify the discussion, we consider a 1-dimensional {\em nominal} model of the multirotor's altitude dynamics, consisting of a point mass model having altitude and its derivative, $[p_z, \dot{p}_z]^T$, as states, mass $m$, and total thrust, $T$, as input:
\begin{equation} \label{eq:drone_nominal}
\begin{bmatrix}{}
\dot{p}_z \\ \ddot{p}_z
\end{bmatrix} = \begin{bmatrix} 0 & 1\\ 0 & 0\end{bmatrix} \begin{bmatrix} p_z \\ \dot{p_z} \end{bmatrix} + \begin{bmatrix}0 \\ 1/m \end{bmatrix} T .
\end{equation}
Using this model we design a nominal MPC as described in Section \ref{sec:MPC_details} with the goal of reaching a fixed point of \SI{0.05}{\metre} above ground at zero velocity. 

A nominal MPC stabilizes the drone to a fixed point, but uses more control effort and time to reach that point as a result of its simplified model. Importantly, the nominal dynamics model does not capture the ground effect. Our goal is to iteratively learn a better dynamics model (and associated MPC) that will improve speed and tracking performance in both the air and near-ground regimes.

\subsection{Implementation and Experimental Details} \label{sec:implementation_details}
Our experiments use the \textit{Intel Aero RTF} Drone.
Drone position is measured using an \textit{OptiTrack} motion capture system and is fused with the drone's IMU (stock PX4 v1.8) to estimate the state. The diffeomorphism, $h$, is parameterized by a neural network and implemented with \textit{PyTorch} \cite{Paszke}, and the KEEDMD regression is implemented with elastic net regularization in \textit{Scikit-learn} \cite{PedregosaFABIANPEDREGOSA2011}. A dense form MPC-controller is implemented in Python using the QP solver \textit{OSQP} \cite{Stellato2018}, and commands are sent to the PX4 flight controller via \textit{ROS}. All computation for learning and control is done on board the drone. Each neural network and MPC evaluation takes 5 ms, limiting us to 5 episodes as update rates below 60 hz lead to poor performance on our hardware. 
The experiment's key parameters are summarized in Table \ref{tab:experiment_params}. 
\begin{table}[b]
\vspace{-0.3cm}
  \centering
  \caption{Experiment Parameters}
    \begin{tabular}{p{2.6cm}rp{2.6cm}r}
    \toprule
    State error penalty, $Q$ & $[10,0.1]$ & Min thrust, $u_{\min}$ & 0.3 \\
    Control penalty, $R$ & 1 & Max thrust, $u_{\max}$ & 0.8\\
    Min altitude, $x_{\min}$ & \SI{0.05}{\metre} & Hover thrust, $u_{\text{hover}}$ & 0.66\\
    \bottomrule
    \end{tabular}%
  \label{tab:experiment_params}%
  \vspace{-0.2cm}
\end{table}%

We execute Algorithm 2 as discussed in Section \ref{sec:EpisodicKEEDMD} on the drone for three episodes in each campaign.  Each episode starts with 3 repetitions of the following: (1) the drone takes off and moves to an initial point under PX4 control; (2) the lifted controller takes over to stabilize the fixed point and hovers at that point for a second. After 3 repetitions, the drone lands under lifted control, fits the diffeomorphism and KEEDMD models, and repeats the episode. An additional landing sequence is executed to evaluate the performance of the current episode controller.

\subsection{Results and Discussion} 
\begin{figure}[t] 
    \centering
    \includegraphics[width=0.48\textwidth]{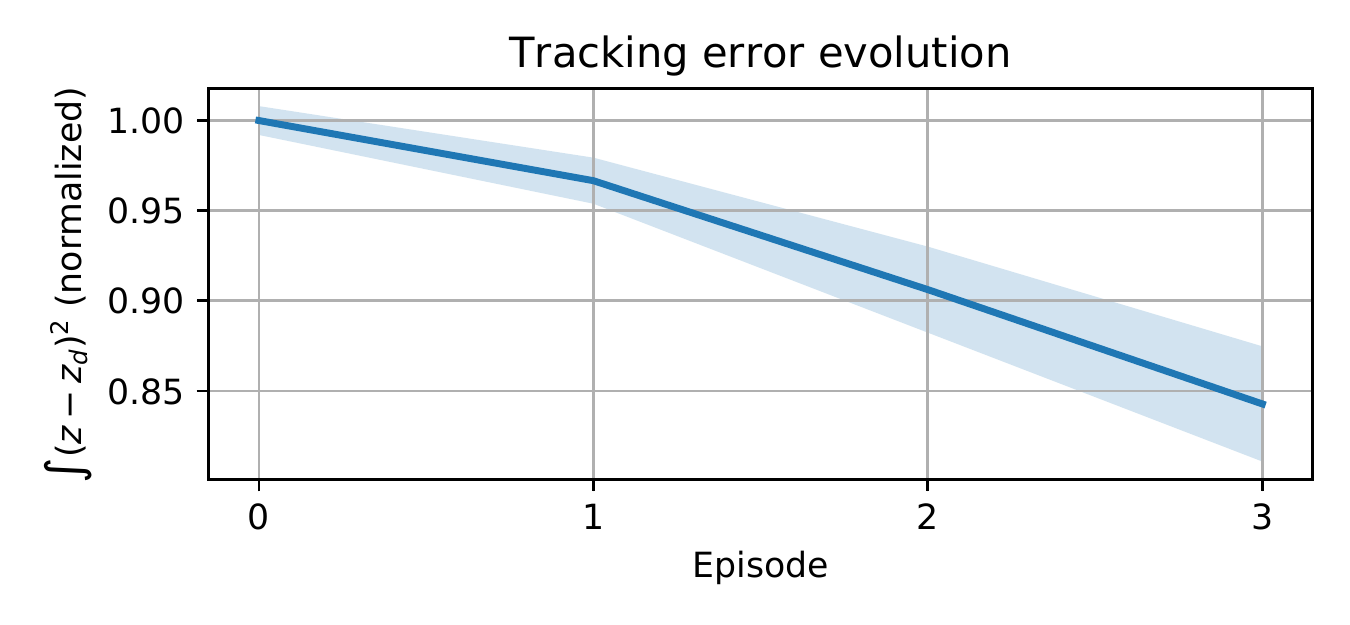}
    \vskip -0.15 true in
    \caption{Mean $\pm$ 1 standard deviation of tracking performance  after each episode over 5 independent campaigns.}
    \label{fig:ep_summary}
\end{figure}

Figure \ref{fig:ep_evolution} depicts the drone's trajectory and control effort under the nominal controller (Episode 0), and then final landing for three episodes of a single learning campaign. Episode 0 represents the nominal performance before learning, while episodes 1-3 show the learning effect. 
Tracking error is reduced by 19.3 percent by the end of the last episode while the total control effort increases 4.5 percent as a consequence of the chosen MPC penalty matrices. Importantly, the thrust constraint is rigorously satisfied, and this constraint is active for longer duration. As the system learns more accurate dynamic models, it relies more on the open-loop bang-bang characteristic, as would be expected from an optimal solution, and less from closed loop control. Less control effort is needed towards the end of the trajectory, indicating that our methodology captures the ground effect.
The mean and standard deviation of five independent learning campaigns are reported in Fig. \ref{fig:ep_summary}. The tracking performance improves in every episode. Furthermore, the methodology has low variance between campaigns. 

\section{Conclusions and Future Work}
This paper presented a novel episodic method based on a Koopman eigenfunction framework to learn a robotic system's nonlinear dynamics, and learn a near optimal control strategy for given tasks. By using a Koopman approach, we are able to implement a real-time MPC framework for optimal system control during the learning process. The approach improves performance as it gathers more data, augmenting the controller to avoid constant actuation matrix limitations, while respecting state and control input bounds. A current limitation is the addition of a controller in each episode leading to prohibitive computational complexity as the number of episodes grows. Current work is addressing this by consolidating the controllers into a finite set, and by optimally choosing when to switch to the next episode, reducing the number of episodes needed.


\section*{Acknowledgement}
This work has been supported in part by Raytheon Company and the DARPA Physics-infused AI program. The first author is grateful for the support of Aker Scholarship Foundation. The authors would like to thank Igor Mezic, Ryan Mohr, and Maria Fonoberova for helpful discussions.

\bibliography{references,references_to_add} 
\bibliographystyle{ieeetr}
\end{document}